\documentclass[12pt]{spieman}  
\usepackage{amsmath,amsfonts,amssymb}
\usepackage{graphicx}
\usepackage{setspace}
\usepackage{tocloft}
\usepackage{algorithm}
\usepackage{algorithmic}

\title{Edge-masked CT image reconstruction from limited data}

\author[a,*]{Victor Churchill}
\author[a]{Anne Gelb}
\affil[a]{Dartmouth College, Department of Mathematics, 27 N. Main St., Hanover, NH, USA, 03755}

\cftpagenumbersoff{figure}
\cftpagenumbersoff{table}
\begin{document} 
\maketitle
\begin{abstract}
This paper presents an iterative inversion algorithm for computed tomography image reconstruction that performs well in terms of accuracy and speed using limited data. The computational method combines an image domain technique and statistical reconstruction by using an initial filtered back projection reconstruction to create a binary edge mask, which is then used in an $\ell_2$-regularized reconstruction. Both theoretical and empirical results are offered to support the algorithm. While in this paper a simple forward model is used and physical edges are used as the sparse feature, the proposed method is flexible and can accommodate any forward model and sparsifying transform.
\end{abstract}
{\noindent \footnotesize\textbf{*}Victor Churchill,  \linkable{Victor.A.Churchill.GR@dartmouth.edu} }

\begin{spacing}{1}   

\section{Introduction}
\label{sect:intro}  




One goal in computed tomography (CT) imaging is to reduce dose while maintaining image quality and reconstruction speed. When collecting less data to reduce dose, direct inversion techniques such as filtered back projection (FBP) produce artifacts. Iterative inversion techniques, like those arising from compressed sensing, have shown that \emph{a priori} information about physical edges in an image can be extremely useful for reconstruction from limited data. For example, assuming edges are sparse in a CT image, one can use total variation (TV) regularization \cite{rudin1992nonlinear}, which solves
\begin{align}\label{eq:l1}
\arg\min_\mathbf{u} \left\{ ||\mathbf{R}\mathbf{u} - \mathbf{s}||_2^2 + \lambda ||\mathbf{D}\mathbf{u}||_1\right\}.
\end{align}
Here $\mathbf{R}$ is the CT forward operator\footnote{The Radon transform is used for data collection and the inverse Radon transform with a Ram-Lak filter for FBP.}, $\mathbf{s}$ is the collected sinogram data, $\mathbf{D}$ is the sparsifying transform\footnote{The anisotropic 2D TV transform defined by $\mathbf{Du} = \sum_{i,j} |\mathbf{u}_{i+1,j}-\mathbf{u}_{i,j}|+|\mathbf{u}_{i,j+1}-\mathbf{u}_{i,j}|$ is used.}, and $\lambda$ is the regularization parameter that balances fidelity, edge sparsity, and noise reduction. In many cases, Eq. (\ref{eq:l1}) is able to reconstruct images of similar quality to FBP using less data. The drawback is the reconstruction speed. In general there is no direct solution to Eq. (\ref{eq:l1}), so an iterative convex optimization method must be used. In this paper, a more direct computational method is proposed, similar to the edge-adaptive $\ell_2$-regularized least squares scheme \cite{churchill2018edge} that was formulated for recovery from non-uniform Fourier data. The method combines an image domain technique with statistical reconstruction. First, the resulting image from an initial FBP reconstruction is used to create an edge mask. Then, an iterative $\ell_2$-regularized reconstruction is performed using the edge mask to regularize away from edges. In what follows, theoretical and empirical evidence is offered to show that with an accurate edge detection, this method has the potential to produce near-perfect reconstructions with extremely limited data at a lower computational cost. In addition, recognizing that the ideal sparse feature for CT images may not be physical edges in the TV domain, this algorithm is flexible and can in fact accommodate a ``wavelet edge" or any another non-physical ``edge'', where the definition is expanded to mean a nonzero location in a sparse domain.
 
\section{Algorithm}\label{sec:methods}
First, the importance of edge locations in image reconstruction is demonstrated by proving that given exact edge locations in the sparsity domain, Eq. (\ref{eq:nonoise}) returns a unique edge-sparse solution if there is one.

\vspace{2ex}\noindent{\footnotesize\textbf{Theorem 1.} Assume there is a unique solution $\mathbf{x}$ to $\mathbf{R}\mathbf{u} =\mathbf{s}$ with $||\mathbf{D}\mathbf{u}||_0\le K$ where $||\cdot||_0$ counts the number of nonzero elements and $K>0$ is a constant. Let $\mathbf{u}^*$ be the solution to the problem
\begin{align}\label{eq:nonoise}
\arg\min_\mathbf{u} ||\mathbf{M}\mathbf{D}\mathbf{u}||_2^2 \quad \text{subject to}\quad \mathbf{R}\mathbf{u} = \mathbf{s},
\end{align}
where $\mathbf{M}$ is a diagonal mask matrix with diagonal element $ 1$ if $(\mathbf{D}\mathbf{x})_i = 0$ and $0$ otherwise. Then $\mathbf{u}^*=\mathbf{x}$.

\noindent\textbf{Proof.} We have that
\begin{align}
||\mathbf{M}\mathbf{D}\mathbf{u}||_2^2 = \sum_{i:\left(\mathbf{D}\mathbf{x}\right)_i=0} \left(\mathbf{D}\mathbf{u}\right)_i^2\ge0,
\end{align}
which is minimized (at zero) for $\mathbf{u}=\mathbf{x}$, since we know $\mathbf{R}\mathbf{x} =\mathbf{s}$. Since $\mathbf{x}$ is the unique solution with $||\mathbf{D}\mathbf{u}||_0\le K$, it must be that $\mathbf{u}^*=\mathbf{x}$.$\quad\square$}

It is critical to note that this theorem would also hold using the $\ell_1$ norm instead of the squared $\ell_2$ norm in Eq. (\ref{eq:nonoise}). That being said, $\ell_2$ is appropriate here because $\mathbf{MDu}$ is not sparse but zero, therefore doesn't require a sparsity-encouraging norm like $\ell_1$ to regularize. Therefore, using $\ell_2$ gives the same result while being more computationally efficient.

\begin{figure}[h]
\begin{center}
\begin{tabular}{c}
\includegraphics[height=2.5cm]{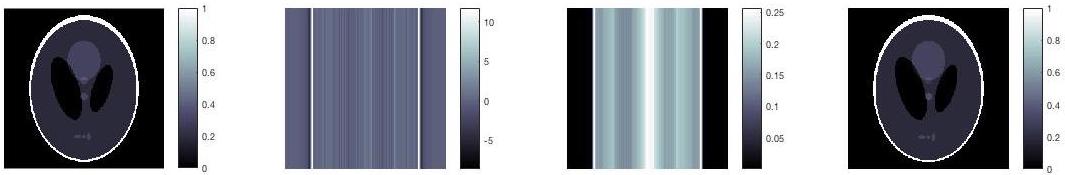}
\end{tabular}
\end{center}
\caption 
{ \label{fig:edgegiven}
Comparison of reconstructions from a single-view sinogram. (left) Shepp-Logan phantom. (left center) FBP reconstruction. (right center) TV-regularized reconstruction from Eq. (\ref{eq:l1}). (right) edge-masked reconstruction given exact edge locations from Eq. (\ref{eq:nonoise}). The right image has relative error $0.0081$.} 
\end{figure}

Figure \ref{fig:edgegiven} shows an example of this theorem in practice compared with FBP and Eq. (\ref{eq:l1}). In this experiment, the exact edge locations are known so the mask $\mathbf{M}$ is prescribed as in Theorem 1 and reconstructed via Eq. (\ref{eq:nonoise}). However, only a single-view sinogram (one projection line) is used. Still, the reconstruction is near-perfect. This example is not intended to serve as a rigorous lower bound for the amount of data required to reconstruct given the exact edge locations, but rather it simply demonstrates the potential of Eq. (\ref{eq:nonoise}) -- that with accurate edge detection, masking provides highly accurate results even with a single-view sinogram.

Theorem 1 does not tell us \emph{how} to collect edge information from CT sinogram data (especially sparse-view data), however. Indeed this is the limiting factor in the method at this point. In this proof-of-concept paper, the edge map is obtained by transforming the FBP image, which is efficient to compute, into the edge domain by the sparsity transform $\mathbf{D}$, and thresholding. There are likely more accurate options. For example, it is possible to design a filter in the FBP process that will suppress most information besides edges. Once the edge map is obtained, the image is constructed by solving the noise-present variant of Eq. (\ref{eq:nonoise}). The full method is given by Algorithm \ref{alg:em}.

\begin{algorithm}
\caption{Edge-masked $\ell_2$-regularized least squares CT image reconstruction}
\label{alg:em}
\begin{algorithmic}[1]
\STATE Collect sinogram, $\mathbf{s}$, and construct an image, $\mathbf{x}_{FBP}$, via FBP.
\STATE Obtain an approximate edge function $\mathbf{y}=\mathbf{Dx}_{FBP}$.
\STATE Define a diagonal mask matrix, $\mathbf{M}$, by
\begin{align}
\mathbf{M}_{ii} = \left\{\begin{matrix}
1 & |\mathbf{y}_i|<\tau\\
0 & |\mathbf{y}_i|\ge\tau
\end{matrix}\right.,
\end{align}
where $\tau$ is a user-defined threshold. 
\STATE Solve
\begin{align}\label{eq:l2}
\arg\min_\mathbf{u}\left\{ ||\mathbf{R}\mathbf{u}-\mathbf{s}||_2^2+\lambda||\mathbf{M}\mathbf{D}\mathbf{u}||_2^2\right\}.
\end{align}
\end{algorithmic}
\end{algorithm}

The newly introduced threshold parameter $\tau$ is of particular importance. As in Theorem 1, the success of the algorithm hinges on having an accurate edge map. Empirically, choosing a higher value $\tau$ may remove some smaller edges but overall yields better results than a less-restrictive edge map, with lower value $\tau$, that might include some false edges created by noise or artifacts from the limited number of views and the FBP process (see Figures \ref{fig:edges} and \ref{fig:edge} later). The regularization parameter, $\lambda$, on the other hand was empirically shown to be not as critical when edge-masking was used \cite{churchill2018edge}.

\section{Results}\label{sec:results}
\begin{figure}[h]
\begin{center}
\begin{tabular}{c}
\includegraphics[height=2.5cm]{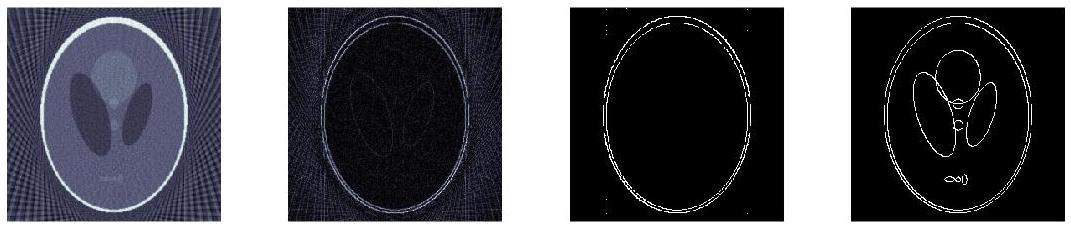}
\end{tabular}
\end{center}
\caption 
{ \label{fig:edges}
Edge mask creation. (left) 45-view FBP image. (left center) FBP image transformed to edge domain. (right center) Thresholded binary approximate edge mask with $\tau=0.3$. (right) True binary edge mask.}
\end{figure}

\begin{figure}[h]
\begin{center}
\begin{tabular}{c}
\includegraphics[height=2.5cm]{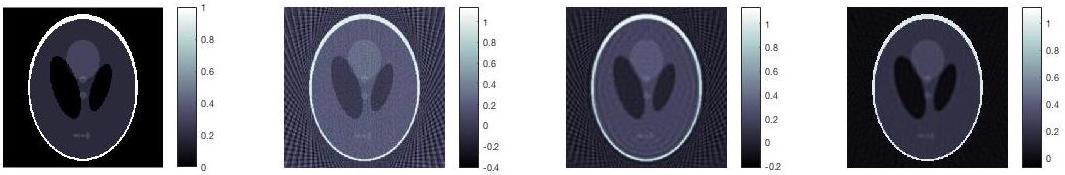}
\end{tabular}
\end{center}
\caption 
{ \label{fig:edge}
Comparison of reconstructions from a $45$-view sinogram. (left) Shepp-Logan phantom. (left center) FBP reconstruction (relative error $0.3783$). (right center) TV-regularized reconstruction from Eq. (\ref{eq:l1}) using $\lambda=0.01$ (relative error $0.3011$). (right) edge-masked reconstruction via Algorithm \ref{alg:em} using $\tau=0.3$ and $\lambda=0.1$ (relative error $0.0888$).}
\end{figure}

As an example, an image is reconstructed from a 45-view sinogram. Figure \ref{fig:edges} shows the edge mask creation process. In this example, the edge map to be used in Algorithm \ref{alg:em} only includes the outer edges of the phantom. Despite this edge mismatch, in Figure \ref{fig:edge} the image is compared with FBP and Eq. (\ref{eq:l1}) and still achieves the most accurate result. This demonstrates the robustness of the algorithm to perform well even when not all edges are present in the mask.

In these results the conjugate gradient algorithm is used to solve Eq. (\ref{eq:l2}). To solve Eq. (\ref{eq:l1}) the Split Bregman method \cite{goldstein2009split}, which requires multiple conjugate gradient iterations, is used. It was set to use $10$ iterations, such that the solve takes about $10$ times longer than Algorithm \ref{alg:em}, but more could be necessary. Hence Algorithm \ref{alg:em} has a distinct advantage in terms of reconstruction speed.

\section{Conclusion}

This paper offers theoretical evidence to explain the potential importance of edge locations in CT image reconstruction. A practical scheme is then developed to utilize this theory. Finally, empirical evidence is presented for using edge-masked regularization to improve accuracy and speed in CT image reconstruction using limited data in order to reduce dose.

There is still much to explore. First and foremost, realizing that the optimal sparse feature for many CT images may not be physical edges, this algorithm is flexible and can accommodate any non-physical ``edge'', where the definition is expanded to mean a nonzero location in a sparse domain. Hence we hope to explicitly explore other sparsifying transforms used in CT. In addition, we hope to test our method on more realistic CT forward models. Finally, we believe that because of its reconstruction speed our method has excellent potential for 3D reconstruction.

\bibliography{report}   
\bibliographystyle{spiejour}   

\end{spacing}
\end{document}